# From Static to Dynamic: Exploring Temporal Networks in Systems Biology

Abir Khazaal[1,2,3], Fatemeh Vafaee[1,2,3,*]

[1] School of Biotechnology and Biomedical Sciences, Faculty of Science, University of New South Wales, Sydney, NSW 2052, Australia

[2] UNSW Biomedical AI, University of New South Wales, Sydney NSW 2052, Australia

[3] UNSW AI Institute, University of New South Wales, Sydney NSW 2052, Australia

[*] Corresponding author:

A/ Prof. Fatemeh Vafaee

T: +61 (2) 9065 2699

E:  f.vafaee@unsw.edu.au

# Abstract

Network science has become an essential interdisciplinary tool for understanding complex biological systems. However, because these systems undergo continuous, often stimulus-driven changes in both structure and function, traditional static network approaches frequently fall short in capturing their dynamic nature. Dynamic network analysis (DNA) addresses this limitation and offers a powerful framework to investigate these evolving relationships. This work focuses on temporal networks, a central paradigm within DNA, as an effective approach for modelling time-resolved changes in biological systems. While DNA has gained traction in domains like social and communication sciences, its integration in biology has been more gradual, hindered by data limitations and the need for domain-specific adaptations. Aimed at supporting researchers, particularly those new to the field, the review offers **an** integrative overview of the diverse and multidisciplinary landscape of DNA, with a focus on temporal networks in systems biology. I begin by clarifying foundational terminology and concepts, then present a multi-scale perspective spanning microscale (nodes and edges), mesoscale (motifs and communities), and macroscale (global topology) analyses. Finally, I explore analytical strategies and computational tools suited to various research objectives, including methods for detecting structural shifts, assessing network similarity, tracking module evolution, and predictive modelling of future network states.

# Introduction

To gain an understanding of a large, complex system, we need to look at the bigger picture instead of focusing on the finer details. Consider the cell at the fundamental level - a densely organised space with RNA, proteins, and various molecules. These cellular constituents engage in numerous highly regulated biochemical reactions. It is challenging to monitor these reactions simultaneously in detail, even at the single-cell level. In addition, we are talking about intracellular reactions and haven't even considered the intercellular interactions, which further demonstrate how intricate the system is [1]. Complex systems are pervasive in our world, spanning from the World Wide Web to neural networks in the brain, from social relationships to atomic-level interactions in folded protein structures involving amino acids [2–5]. Such systems are usually represented through networks, modelled as mathematical graphs. A network consists of a set of nodes representing the system's entities - also known as points, vertices or actors in the field of social science, and a set of edges (links) symbolising the connections between them [6,7]. Since the early 20$^{th}$ century, the study of networks, or *network science*, emerged as a cross-disciplinary field, combining approaches from graph theory, data mining, statistical inference and information visualisation methods [8,9]. It is now recognised as a fundamental framework for describing the dynamics of complex systems, allowing us to understand their behavior through a holistic view rather than concentrating on individual dynamics [10,11]. A concrete example of this trend is the notable upswing in reviews on networks across various fields [7,12–14], such as studies leveraging protein-protein interaction (PPI) networks in biology [15–18]. This surge is attributed not only to the progress in tools and techniques but also to the exponential growth of big digitised data, facilitated by advancements in computer power [7,19]. For example, introducing high-throughput technologies has enabled researchers to generate vast amounts of available meta-omics data, underscoring these technological improvements' profound impact [20–22].

Network approaches shift the emphasis from merely examining the attributes of individual units to analysing the relationships and patterns that emerge from their interactions. While traditional network analysis has proven to be a powerful tool, its effectiveness can be leveraged by incorporating additional layers of detail [23–25]. For instance, edges can be given weights to reflect the strength of connections, leading to weighted networks, or assigned orientations to differentiate between directed and undirected networks [26,27]. A significant advancement in network studies has been the integration of time as a crucial dimension [28–30]. Real-world networks are rarely static; they are dynamic, continuously evolving over time or in response to perturbations [31–34]. Despite this understanding, much of the earlier literature concentrated on cross-sectional studies. This focus was primarily due to the limited availability of longitudinal data and the prevailing belief that time as a dimension should be suppressed or simplified to ease the analysis [35–37]. The emergence of *dynamic network analysis* (*DNA*) has transformed the field, offering researchers powerful methodologies to explore the temporal and changing aspects of evolving relationships [38–40].

Although *DNA* has gained considerable traction in fields such as social sciences (e.g. social networks), transportation networks (e.g. airline routes) or information networks (e.g. hyperlinks between webpages), its adoption in biological sciences has been slower. This delay is partly because biological data has only become more accessible in recent years [41]. Additionally, when available, such data often has lower temporal resolution due to experimental and financial constraints, unlike other fields where high-resolution data, such as in telecommunication networks (e.g. email and phone data), is more readily abundant [42,43]. Recent studies, nevertheless, are beginning to bridge this gap.

In addition to data constraints, the distinctive nature of biological networks and the questions they address further complicate the direct application of established *DNA* concepts. Researchers often had to substantially adapt existing network analysis methods, originally developed for other disciplines, to meet the unique challenges presented by biological data. For instance, in social sciences, concepts like *information and influence propagation* are crucial for understanding the spread of ideas or behaviors within and across networks. In biology, these concepts take on different forms. They typically manifest through mechanisms such as signaling pathways, cell-cell communication, or gene regulatory networks (GRNs) [44,45]. Unlike the diffuse and large-scale propagation of ideas in social sciences, biological information transfer is highly specific, occurring at the cellular or molecular level. For instance, cell-cell communication involves the transmission of signals from one cell to another, often through chemical messengers like hormones or neurotransmitters. These signals coordinate cellular responses and maintain homeostasis [46,47]. GRNs, on the other hand, encompass complex interactions between genes, proteins, and other molecules within the cell. These networks control cellular processes, ensuring that genes are expressed at the right time and in the right amounts [48,49]. While both fields deal with the propagation of information, biology focuses on precise, tightly regulated communication at a cellular or molecular level, in contrast to the broader, societal level seen in social sciences.

In this review, I aim to explore the interdisciplinary field of *DNA*, with particular attention to its applications in biological networks. The term "biological networks" broadly refers to the complex interactions within biological systems, such as those between immune cells, metabolites within cells, and genes and proteins in the nucleus. While our review covers biological networks of all kinds, the specific adaptation of methods depends on the data and networks in question. I begin by exploring the concept of dynamic networks and discussing the various terminologies under which they appear in the literature. Next, I examine the characteristics of temporal networks at different scales: starting at the microscale level by analysing nodes and edges, progressing to the mesoscale level by considering network motifs and communities, and finally, expanding to the global scale by assessing the network as a whole. In the third section, I present a range of analytical approaches used to study temporal networks, including both dedicated and repurposed methods. Lastly, the review concludes with a synthesis of insights and final reflections on future directions in this evolving field.

## Dynamic networks

As mentioned earlier, networks change and evolve either due to the passing of time or in response to stimuli, phenomena or perturbations. These changes can manifest in three distinct aspects. The first aspect involves changes to the nodes themselves, even when the edges remain constant. For example, PPI networks in the MAP kinase signal transduction pathway play a crucial role in controlling processes such as cell growth, differentiation, and immune-related inflammatory responses [50–52]. In cancer, for instance, an altered form of protein, such as serine/threonine-protein kinase B-Raf (BRAF), may replace its wild-type counterpart within the signaling network. Although the mutated BRAF continues to interact with downstream proteins, like MEK, thus preserving the same signal relay (edge) in the network, the node (the specific protein form) has changed. This alteration can impact how the signaling network operates [53–55]. A second aspect of changes involves the edges rather than the nodes. Many studies on Mendelian disorders - diseases caused by mutations in a single gene - have shown that most missense mutations disrupt either all or a subset of interactions within PPI networks [56]. Some researchers even refer to these as "edgetic" mutations or perturbations [57,58]. The third aspect of network changes occurs when both nodes and edges are affected simultaneously. A prime example of this can be found in adaptive immunity, the immune system's specific defence mechanism. In this context, naive T cells (CD4+) can differentiate into effector T cells or regulatory T cells, and this differentiation is governed by distinct transcription factors (nodes), T-bet and Foxp3 respectively, each initiating different cascades of gene interactions (edges) [59,60]. Analysing dynamic networks where both nodes and edges change presents significant challenges. However, ongoing advancements in the field will likely continue to address these complexities, offering deeper insights into the intricate involved [61].

Although the interdisciplinary nature of DNA can be seen as a strength, the variety of frameworks and inconsistent terminology in the literature can pose difficulties, particularly for those new to the field. For example, dynamic networks can be referred to as evolving networks [62,63], evolving graphs [64,65], temporal networks [66,67], temporal graphs [68,69], dynamic graphs [70,71], time-varying networks [72,73], time-varying graphs [74,75], multislice networks [76,77] and systems biology dynamical models [78,79]. Similarly, DNA as a methodology is found under terms such as evolutionary network analysis [80,81], dynamic social network analysis [82,83], network dynamics [84], network alignment [85], graph matching networks [86,87], network similarity methods [88], network rewiring [89], multiplex networks [90] and multilayer networks [91]. While some of the terms mentioned may not be directly associated with dynamic networks, such as graph matching networks, which are used to compute similarity between graph-structured objects, they can still be applicable in this context [87,92]. The goal is not to provide an exhaustive review of the dynamic network literature, as several excellent reviews already exist [11,28,30,37,93]. Instead, I seek to offer a broad overview for the reader, focusing on bridging the gap between dynamic network concepts and their application in the biological sector. Moving forward, we will use temporal networks (TNs) as a prime example to illustrate the types of analyses that can be performed. By definition, TNs represent systems where the network structure changes over time [94]. It is important to emphasise, though, that this approach is broadly applicable to dynamic networks as a whole.

# Temporal Networks Characteristics

Dynamic networks, particularly Temporal Networks (TNs), differ significantly from static ones, presenting distinct analytical challenges. TNs generally tend to be larger in scale, often more complex and information-rich due to the added dimension of time. Network sizes and node counts can fluctuate, and the level of edge activity varies across different time points. In the case of weighted networks, for example, an edge may persist across several time points, but its weight or interaction intensity may change. TNs can be analysed at multiple levels to understand these evolving structures: micro-scale, meso-scale, and global-scale. Each level offers unique insights and requires specific tools and approaches tailored to the complexities of network dynamics, ensuring a more comprehensive understanding of the system's evolving behavior.

**Micro-scale level**

*Nodes*

At the micro-scale level of network analysis, several key measurements and properties are used to evaluate the role and behavior of individual nodes. A node's *degree* refers to the number of edges connected to it. In GRNs, this represents the number of genes interacting with a given gene (node). Generally, a higher *degree* indicates greater node importance [95]. In TNs, a node's *degree* can vary, allowing researchers to track changes in node connectivity over time and observe whether a node is gaining or losing influence. Two categories of "influential" nodes can be distinguished: those essential within a specific network at a given time point and those that remain indispensable across multiple networks over different time points [96]. That said, in TNs, traditional degree-based metrics lose some relevance when identifying "influential" nodes because network sizes (i.e., the number of nodes) often fluctuate, making direct comparisons of node *degrees* across different networks impractical.

To address this, alternative measures have been proposed to identify central nodes more accurately [97,98].

Another important class of metrics is *centrality*, which assesses a node's importance within the network. Several types of centralities can be measured. *Degree centrality*, closely related to the concept of *degree*, is calculated by dividing the total number of edges connected to a node by the maximum possible number of connections, normalising the value to allow for meaningful comparisons across networks [99]. This measure identifies *hubs* - nodes that are critical for maintaining network stability. *Degree centrality* is particularly important because it reflects how robust a network is through the presence of these *hubs* [100]. In Biology, the "centrality-lethality" rule further underscores the significance of *hubs* in maintaining the stability of PPI networks [101,102]. *Betweenness centrality* is another well-studied measure that defines a node's importance by evaluating how often it lies on the shortest paths between other nodes [101,103]. The shortest path (also known as geodesic distance) is the path with the fewest edges between two nodes [104–106].

Beyond traditional centrality measures, additional network metrics have been introduced to identify a minimal set of key nodes that exert full control over the network, offering significant potential in identifying disease-associated genes and drug targets [107,108]. Unlike centrality-based approaches that assign importance based on connectivity, these methods focus on structural controllability by determining the smallest subset of nodes (minimum dominating set) required to influence the entire network [109]. In breast cancer, drug target genes have been identified using an optimisation framework designed to capture dominant key regulatory nodes, demonstrating the effectiveness of such approaches in uncovering critical therapeutic targets [110,111].

In TNs, these centrality measures adapt to account for the dynamic nature of interactions. For instance, *temporal betweenness centrality* considers the shortest paths and respects the timing of interactions, ensuring that the paths follow a logical sequence in time (time-respecting paths) [112,113]. This metric is particularly useful in fields like epidemiology, where it is applied to study processes such as the spread of infectious diseases [114,115]. The applications of these metrics in biological networks require specific interpretation. *Temporal degree centrality* can be used to identify key transcription factors (*hubs*) that become central at different stages of cell differentiation, highlighting their roles in driving developmental processes [116]. Similarly, this measure has been applied in PPI networks to study the dynamic behavior of genes at various stages of diseases, such as cancer. For example, in pancreatic ductal adenocarcinoma (PDAC), temporal centrality measures, including *betweenness* and *degree centrality*, have been used to identify key genes that play crucial roles in different stages of cancer progression. Genes such as NDC80, KIF2C, and ITGB1 were identified as having high centrality in the early stages of PDAC, while other genes like ITGA4 and ITGB4 became more central in the advanced stages, driving the network's stability and progression [117]. These insights would likely have been missed in a static network analysis.

*Edges*

Edges in TNs represent dynamic interactions between nodes, changing over time in activity, strength, or direction. Unlike static networks, where edges remain fixed, TN edges evolve based on temporal factors such as biological processes. Examining these temporal shifts is particularly valuable in understanding regulatory interactions during key biological transitions.

For example, during T-cell activation, cytokine-receptor interactions fluctuate with antigen exposure, shaping immune responses over time [118,119]. Similarly, in GRNs, regulatory interactions between genes emerge or vanish depending on specific conditions or developmental stages. This dynamic nature is crucial for understanding gene activation, suppression, and coordinated expression in processes like stress responses and cell differentiation [120].

A notable example is the PU.1 gene, a lineage-determining transcription factor critical for macrophage and B cell development [121]. PU.1 regulates gene expression by binding to distinct enhancer elements in a cell-specific manner. In macrophages, it interacts with C/EBPs and AP1, whereas in B cells, it associates with factors such as EBF1, E2A, and OCT. These dynamic interactions illustrate how transcription factor networks are highly context-dependent, with PU.1's regulatory edges shifting based on the cellular environment [122]. This highlights the limitations of static networks, which fail to capture such context-driven changes. Similarly, in PPI networks, the interaction between proteins fluctuates in response to environmental stimuli or cellular states [123]. Analysing these time-dependent changes enables researchers to identify critical periods of interaction, contributing to system-level processes such as cell differentiation, immune response, or disease progression [124].

Furthermore, the weight of edges in TNs can reflect the intensity or frequency of interactions, adding another layer of complexity. In GRNs, edge weights may indicate the strength of regulatory influence, while in PPI networks, they often correspond to binding affinity. High-affinity interactions are typically crucial for stable complexes, such as enzyme-substrate binding or receptor-ligand interactions [125]. In contrast, low-affinity interactions often play roles in transient or reversible processes, including signaling pathways or the dynamic assembly of protein complexes [126]. Notably, affinity may increase during stress responses, such as heat shock or infection, and decrease under normal conditions [127,128]. In neurobiological networks, synaptic connections between neurons exhibit changes in synaptic strength (edge weight) due to synaptic plasticity during the learning process [129,130]. Conversely, in neurodegenerative diseases like Alzheimer's, edge weights gradually weaken as synapses are lost [131]. These dynamic shifts in synaptic strength emphasise the role of temporal network analysis in understanding learning, memory, and neuronal transmission efficiency, particularly as they change with age.

Additionally, edges in TNs can display continuous activity, intermittent periods of inactivity, or periodic alternations between activity and inactivity at specific time points or intervals [11,30].

In a temporal gene co-expression network, for example, genes can exhibit similar expression patterns only under specific conditions, such as in response to external stimuli. Such mechanisms have been observed in yeast studies [132]. A related example can be found in immune response networks, where interactions between immune cells—such as T cells and B cells—are only activated during specific phases of infection or inflammation, reflecting the episodic nature of immune responses [133]. Understanding the temporal dynamics of these interactions—when and why they become active or inactive—offers deeper insights into immune signal transduction and gene expression regulation.

Moreover, directed edges often play a more crucial role in TNs than static networks. When analysing connectivity and reachability in directed TNs—sometimes referred to as scheduled networks—the concept of a time-respecting directed path becomes essential [93]. This path consists of edges with time labels that either remain constant or increase, ensuring connections progress in a biologically meaningful temporal sequence [134]. This concept is particularly significant in biological interactions, such as in GRNs, which are inherently directional [135]. Although all cells share the same genetic information, their structural characteristics and function are determined by the specific set of genes activated by other genes, known as regulators, often transcription factors (TFs). This directional and sequential activation allows us to differentiate master regulators – key genes at the top hierarchy that initiate the cascade of gene interactions essential for cell differentiation and developmental transitions [136].

Overall, the temporal dynamics of edges in TNs underscore the complexity of biological interactions, highlighting how connectivity patterns evolve to regulate cellular functions, adaptive responses, and system-level processes.

**Meso-scale level**

If we zoom out a little bit from nodes and edges, at the meso-scale level, we start to analyse interactions and patterns within subgroups of nodes, which could indicate important functional properties. Concepts like subgraphs, graphlets, motifs, cliques, and communities are all related and sometimes used interchangeably, but serve distinct characteristics and properties.

*Subgraphs*

Subgraphs is more of a general term that refers to any subset of a larger graph and is generally used if we want to study specific parts of a network. Often referred to as subnetworks in network science, they consist of a subset of nodes and edges, ranging from a single node to the entire graph [137,138]. Subgraphs can be categorised as induced or non-induced (partial), depending on whether all edges among the chosen nodes are included [139]. They are considered a foundational concept, with some researchers proposing *subgraph centrality* as a measurement to characterise and rank nodes. Interestingly, *subgraph centrality* has been observed to follow a power-law distribution, even in cases where the degree distribution of nodes does not [140,141]. By defining specific criteria or features within subgraphs, we can identify more specialised structures, such as graphlets, motifs, and communities. Figure 1 illustrates how subgraphs and communities represent distinct substructures within a larger network.

*Graphlets*

While subgraphs refer to any subset of a graph, graphlets are specifically small, induced subgraphs consisting of 2-5 connected nodes that are used to uncover structural features of a network [142]. They are often counted to learn features of network structures, describing their "fingerprint" [143]. For instance, graphlets were employed in early analyses of PPI networks, where researchers analysed local structures of nodes using a node's *graphlet degree* rather "as a metric" than relying solely on its *degree*. This approach enabled deeper insights into predicting protein function based on the structural context within the network [144]. Similarly, graphlet analysis has been applied to effective brain networks, revealing differences in local structures between excitatory and inhibitory circuits and offering valuable insights into neural connectivity patterns [145].

*Network motifs*

Network motifs can be defined as over-represented graphlets, observed within a given network more frequently than expected by chance [146,147]. These recurrent patterns of interactions represent functional and organisational importance within networks, underpinning their complexity and efficiency by enabling robust and adaptive behaviors in complex systems [148]. Transcription networks, which consist of interactions between transcription factors and their target genes, provide a prime example of this principle [149]. Motifs initially identified in the transcription network of *Escherichia coli* have been found to share structural and functional similarities with those in diverse species, including yeast [150], plants [151], and humans [152], indicating a fundamental role in gene regulation networks [153]. In directed networks, we can distinguish different types of motifs, such as the feed-forward loop (FFL) [154] and feed-back loop [155] for three-node motifs, as well as bi-fans [156], bi-parallels [157], and other configurations for motifs involving four or more nodes [158–160]. Since motifs have become essential for extracting meaningful insights from large and complex networks, numerous methods and algorithms have been developed to detect, count, and cluster motifs into higher-order structures on a large scale, facilitating efficient network analysis [161].

*Cliques*

Cliques are a special form of motif where every node is connected to all other nodes within the motif [162]. Defined as maximally connected subgraphs, a clique of *n* nodes contains *n-1* edges, representing maximal density in terms of connectivity [163]. Originally prominent in social sciences for studying tightly linked groups (e.g., friendships), cliques have since been adopted across many fields [164,165]. In large networks, the number of cliques grows exponentially, making their identification and enumeration computationally challenging. This has led to the development of Maximal Clique Enumeration (MCE), which focuses on identifying all maximal cliques in a graph [166]. A maximal clique is a complete clique that cannot be extended, in other words, it is not a part of a larger clique [167,168]. MCE has become essential in studying complex networks, with many algorithms implemented to address it [169,170]. In biology, maximal cliques can represent sets of proteins physically or functionally connected in PPI networks or pinpoint groups of genetic loci co-regulating a particular trait or disease phenotype in quantitative trait loci (QTL) mapping [171,172]. Graphlets, motifs, and cliques represent fundamental network substructures that help reveal connectivity patterns at different scales of organisation (Figure 2)

*Communities*

While analysing motifs and cliques provide valuable insights into networks, these rigidly defined structures are insufficient for capturing the complexity of real-world systems. Such networks are often large, intricate and inherently organised into modular structures, requiring a more coarse-grained approach [173]. In this context, analysing modules in network science parallels the study of communities in graph theory, where the goal is to identify structurally cohesive regions within a network [174]. Although lacking a strict formal definition, communities are broadly characterised as large clusters of nodes that exhibit denser internal connections relative to their links with the rest of the network [175,176]. Significant research has been devoted to the field of community discovery, resulting in numerous algorithms and approaches trying to address this problem [177–179]. Among these, modularity-maximising algorithms have emerged as a prominent class of methods, aiming to partition networks in a way that maximises the difference between observed and expected intra-community connections [180–182]. The growing interest in community discovery stems from the fact that these cohesive groups of nodes often correspond to units of functional importance within the network. In complex diseases, disruptions often involve extensive perturbations across gene interaction networks rather than being limited to single gene alterations. Community detection in PPI networks has the potential to unravel these molecular disruptions, providing critical insights into disease pathology and informing therapeutic strategies [183,184]. These insights not only enhance our understanding of disease progression but also facilitate the identification of potential drug targets, paving the way for novel therapeutic strategies [185]. Table 1 provides a comparative overview of these network substructures, highlighting their definitions, purposes, and typical applications.

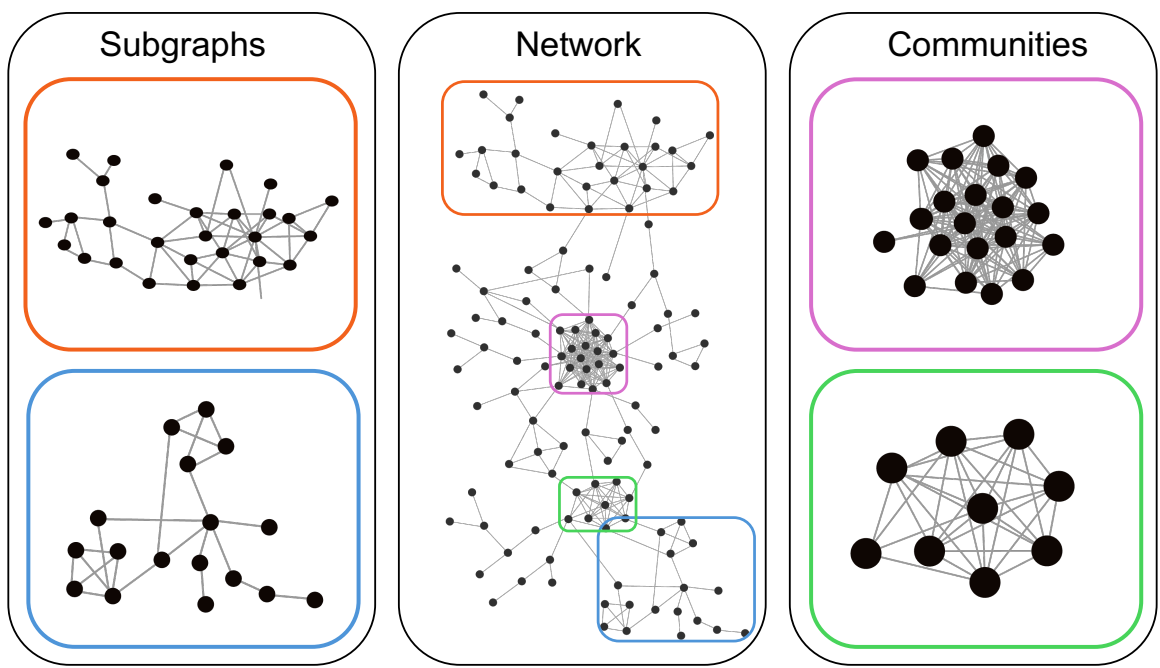

**Figure 1** Network substructures, distinguishing between subgraphs and communities. A network can be analysed at different scales, revealing meaningful substructures. **Middle-Panel**: Network representation shows the overall connectivity of nodes and edges. **Left-Panel**: Subgraphs are smaller, connected subsets of the network, capturing localised interactions or functional modules. Two examples of subgraphs are highlighted (orange and blue). **Right-Panel:** Communities are densely connected groups of nodes within the network, often corresponding to functional or organisational units. Two communities with varying densities are depicted (purple and green).

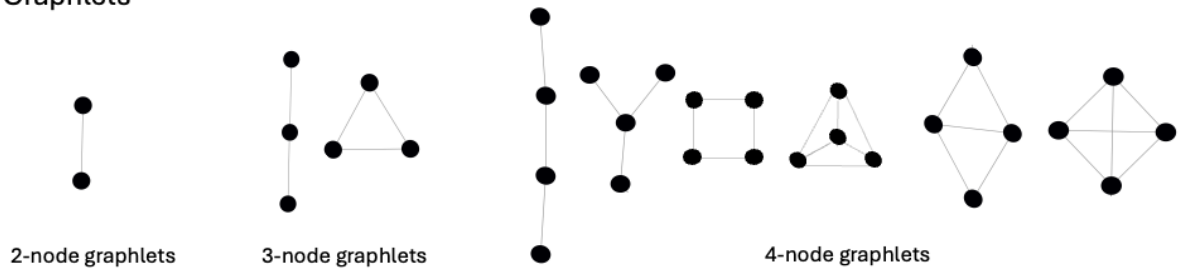
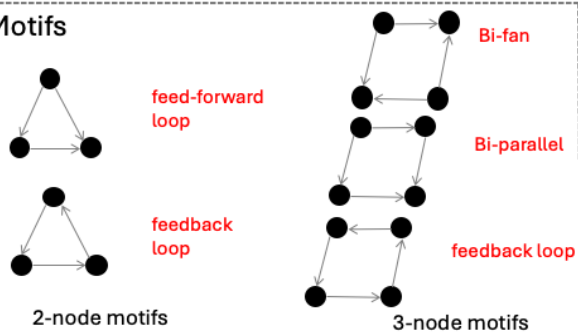
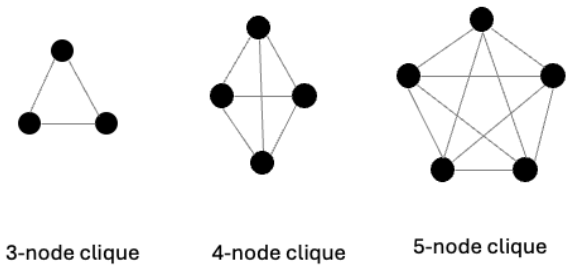

**Figure 2** Graphlets, Motifs, and Cliques: Fundamental Building Blocks of Network Topology. **Top:** Graphlets are small, connected, and non-isomorphic subgraphs that capture local connectivity patterns. Examples of 2-node, 3-node, and various 4-node graphlets are shown. **Bottom left**: Motifs are recurrent, statistically significant subgraph patterns that often have functional implications in biological and technological networks. Common motifs such as feed-forward loops, feedback loops, bi-fan, and bi-parallel structures are depicted. **Bottom right:** Cliques are fully connected subgraphs where each node is directly linked to all others. Examples of 3-node, 4-node, and 5-node cliques illustrate increasing structural complexity.

**Table 1** Summary of key features of mesoscale structures: Subgraphs, Graphlets, Motifs, Cliques, and Communities

| Substructure | Subgraphs | Graphlets | Motifs | Cliques | Communities |
|---|---|---|---|---|---|
| **Definition** | Any subset of nodes and edges | Small, connected subgraphs of fixed size | Statistically significant subgraphs | Maximally connected subgraphs | Densely connected groups of nodes |
| **Size** | Arbitrary | Fixed (e.g., 3–5 nodes) | Small to medium (commonly 3-5, can be up to 10) | Small to medium (3-6 nodes) | Varies, larger than motifs |
| **Purpose** | General study of parts of a graph | Local structural analysis | Identify key functional patterns | Identify strong functional connections | Reveal mesoscale network structure |
| **Key Focus** | Structural flexibility | Standardised small patterns | Statistical overrepresentation | Complete and dense locally connectivity | Dense internal connections |
| **Example** | Subnetwork of metabolic pathways | Triangle, star, or chain | Feed-forward loop motifs in transcription networks | protein complexes in PPI networks | Functional modules in gene regulatory networks |

*Meso-scale structures in TNs*

The dynamic nature of real-world networks adds complexity to the analysis of mesoscale structures, as these features can emerge, persist or dissolve with time. Understanding their behavior in TNs requires novel methods that account for their dynamic properties, going beyond static analyses to explore their temporal evolution.

In biological networks, temporal data resolution is often low, reducing the prevalence of challenges like arbitrary timescale selection – when fixed time windows are imposed – which is a significant issue in other fields [186]. Instead, TNs in biology are frequently represented as snapshots of networks captured at different time points, allowing researchers to track changes in mesoscale features across time [187,188]. While some may argue that temporal data is increasingly available for processes like ordered molecular phases in biological systems, it remains low-resolution compared to fields such as communication networks [189,190]. The concept of temporal motifs – precise time-ordered sequences of interactions (e.g., A → B → C in a feed-forward loop) – is less relevant in biology due to these resolution constraints [191,192].

Instead, the focus shifts to identifying over-represented mesoscale features – such as motifs, cliques, or communities – that persist or appear transiently over time. For example, in PPI networks, persistent cliques may represent stable protein complexes, where groups of proteins interact consistently to form core functional modules, such as those found in the ribosome or spliceosome [193,194]. By contrast, transient cliques often arise from short-lived protein interactions associated with dynamic processes, such as signaling cascades or transient enzyme-substrate interactions, offering insights into regulatory events [193,195].

Similarly, the study of temporal communities provides a powerful framework for understanding the higher-order dynamics of networks. Rather than remaining fixed, these structures evolve, reflecting shifts in biological organisation and function. Communities may grow as new proteins, or molecular entities integrate into an existing module or contract when components become less relevant or disengage. Merging and splitting highlight the modular reorganisation of biological interactions, where functional units combine into larger assemblies or fragment into specialised substructures in response to environmental or regulatory changes. Some communities emerge (birth) as novel interactions form, while others dissolve (death) when their functional relevance declines (Figure 3). Capturing these temporal patterns provides deeper insights into network plasticity, adaptation, and stability, with implications for cellular responses, disease progression, and molecular evolution.

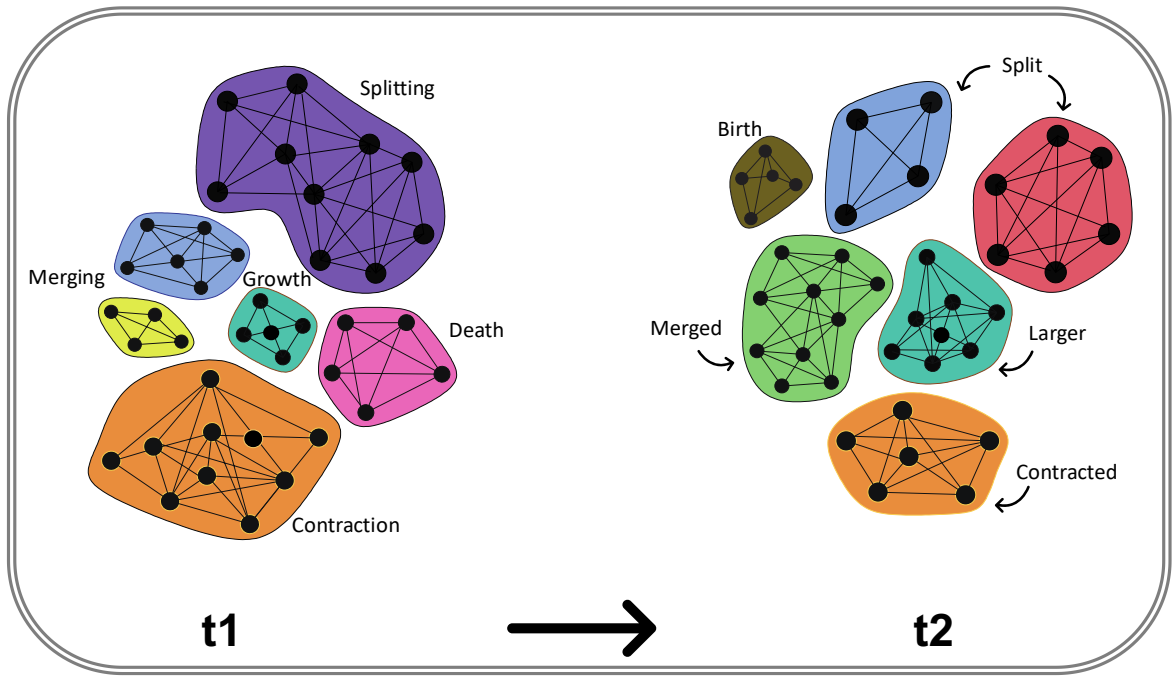

**Figure 3** Community dynamics in temporal networks. Left to right: Communities are shown at two time points $t_1$ and $t_2$, with color denoting identity and transformation over time. The blue and yellow communities at $t_1$ merge into a single green community at $t_2$ (merging), while the purple community splits into red and blue substructures (splitting). The aqua community expands by incorporating new nodes (growth), whereas the orange community contracts (shrinkage). A novel brown community appears at $t_2$, indicating emergence (birth), while the pink community dissolves (death), disappearing entirely by $t_2$. These temporal transitions illustrate how biological modules reconfigure, adapt, and reorganise in response to internal or external pressures, offering insight into the dynamic nature of molecular interaction networks.

Beyond community-level changes, mesoscale organisation also shifts, particularly in the roles of intermodular and intramodular hubs [196]. Intermodular hubs, as the name suggests, are nodes that bridge distinct functional modules. While they can be identified in both static and dynamic networks, tracking their changes over time may reveal additional insights into their regulatory roles. For example, in PPI networks, these hubs may be more vulnerable to mutations, potentially affecting multiple pathways or processes. Conversely, intramodular hubs, which are deeply embedded within specific modules, tend to be essential for the stability and function of core complexes (Figure 4) [4,197]. This principle also applies to disease networks, where tracking the evolution of functional modules over time can reveal condition-specific molecular mechanisms and identify potential therapeutic targets [198].

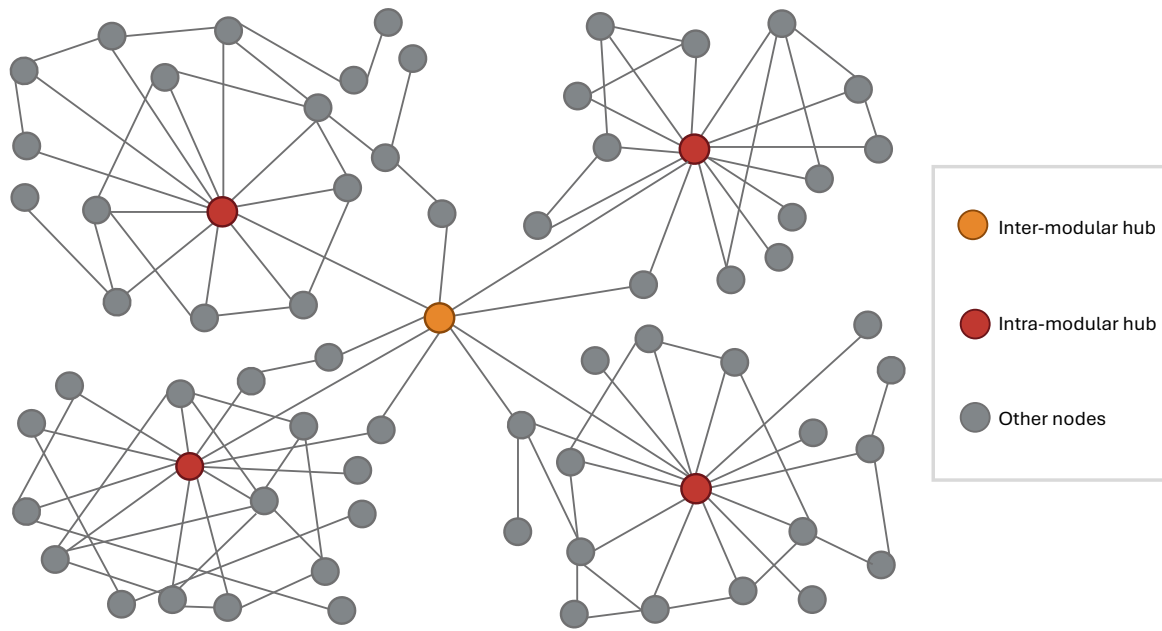

**Figure 4** Intermodular vs. Intramodular Hubs. A network composed of four scale-free modules. Red nodes represent intra-modular hubs, defined as the highest-degree nodes within each module, responsible for maintaining local cohesion and function. The orange node indicates an inter-modular hub connecting to all modules' hubs and other nodes, serving as a bridge between otherwise distinct communities. Such intermodular hubs are important in temporal networks, where their appearance or disappearance across time points can signal critical changes in network-level communication, regulatory coordination, or system-wide transitions. All other nodes are shown in grey.

Accurately capturing the temporal dynamics of these mesoscale structures also requires careful selection of analytical models and methods. These models provide a mathematical and computational framework to track temporal changes, identify emerging patterns and account for network heterogeneity [199,200]. For instance, communities may grow, merge, or dissolve over time, necessitating models that can adapt to such structural transformations [201]. Some models, like first-order Markov chains, assume a memoryless process, where the state of the network at any given time depends only on its immediate past [202]. While this assumption simplifies analysis and reduces computational complexity, it may fail to capture long-range dependencies or cascading effects—particularly important in biological systems. Biological networks are inherently non-linear; for example, a mutation in a gene within a regulatory network can have widespread effects, altering multiple pathways and triggering systemic changes. To address these limitations, higher-order Markov chains extend the framework by allowing the current state to depend on multiple prior states, providing a richer representation of temporal dependencies [203]. However, these models require careful calibration to avoid overfitting (where random noise is mistaken for meaningful structure) and underfitting (where essential temporal patterns are overlooked). Thus, selecting the appropriate analytical model is essential for optimising computational efficiency, maintaining biological realism, and ensuring predictive accuracy in the study of temporal biological networks.

**Macro-scale level**

Network analysis shifts from examining modular structures at the mesoscale level to assessing the entire network's topology and connectivity patterns at the macroscale level. This system-wide perspective encompasses various global metrics that characterise network topology [204]. These metrics provide insights into structural robustness, information flow, and system-wide resilience [12,35].

One fundamental measure is Density, which quantifies how interconnected the network is by considering the ratio of existing edges to the total possible edges [205]. Higher density typically indicates a more cohesive structure, facilitating robust information flow, whereas lower density suggests a sparse network with a fragmented or modular organisation (Figure 5-A) [206,207]. Many clustering algorithms leverage density as a guiding principle to identify functional modules within PPI networks. Typically, they begin with 'seed' proteins and expand clusters by iteratively adding edges while ensuring that the cluster maintains a predefined density threshold [208,209].

Another key metric is the Degree Distribution, which describes the statistical distribution of node degrees—the number of connections each node has [210]. Understanding degree distribution helps determine whether a network exhibits scale-free properties, a defining characteristic of many real-world systems [211,212]. In scale-free networks, the degree distribution follows a power law, where a few highly connected nodes (hubs) coexist with many sparsely connected nodes (Figure 5-B) [213,214]. This structure has profound implications: scale-free networks are highly resilient to random failures – removing many randomly selected nodes often leaves network integrity largely intact [215,216]. However, they are highly susceptible to targeted disruptions—removal of key hub nodes can cause the entire network to collapse, a phenomenon known as attack vulnerability [217,218]. For example, in genetic networks, this structure suggests that certain key genes play central roles in cellular function. While random gene mutations may have minimal effects, losing a critical hub gene could trigger system-wide failures, leading to severe phenotypic consequences [219].

The Clustering Coefficient measures the tendency of nodes to form tightly knit groups or interconnected triads [220]. This metric exists at both local and global levels. The local clustering coefficient evaluates how interconnected a specific node's neighbours are, while the global clustering coefficient—more relevant at the macroscale level—captures the overall prevalence of triangular patterns, reflecting network cohesiveness [221,222]. High global clustering often signifies modular or community-based organisation within a network, whereas low clustering suggests random-like connectivity patterns [207].

A related concept, Modularity, quantifies how well a network can be divided into distinct communities or modules [223]. Networks with high modularity exhibit well-defined clusters with dense intra-community connections and sparse inter-community links. Conversely, networks with low modularity behave more like random graphs, often exhibiting degree distributions that approximate a Poisson distribution (Figure 5-C) [180,224].

Another topological property is the Average Path Length (APL), which represents the mean of the shortest paths between all pairs of nodes in a network [89]. It is recognised as one of the most frequently cited network characteristics and is commonly referred to as Characteristic Path Length in social network studies [225]. This metric provides insight into the efficiency of information, signal, or influence transmission flow across a network. In biological networks, APL – along with other topological measures – can aid in identifying relevant genes in genome-wide association studies (GWAS) [226,227]. For instance, in disease-related research, functionally related genes tend to cluster together within the interactomes, leading to a lower APL compared to randomly selected gene sets [228,229]. Therefore, analysing APL can help identify biologically relevant gene modules and enhance our understanding of disease mechanisms [230].

All these metrics provide valuable insights into the structural and functional organisation of networks. As an example, neuronal networks in the brain have been shown to exhibit small-world property, characterised by a high clustering coefficient (local connectivity within brain regions), short average path length (efficient global communication), and modularity (specialised functional regions with selective long-range connections) [231,232]. These properties collectively ensure computational efficiency and adaptability. Additional macroscale metrics, such as Assortativity (degree correlation), Betweenness Centrality distribution and Network Diameter, and properties such as Resilience, further contribute to characterising network structure; however, they fall beyond the scope of this review [233,234].

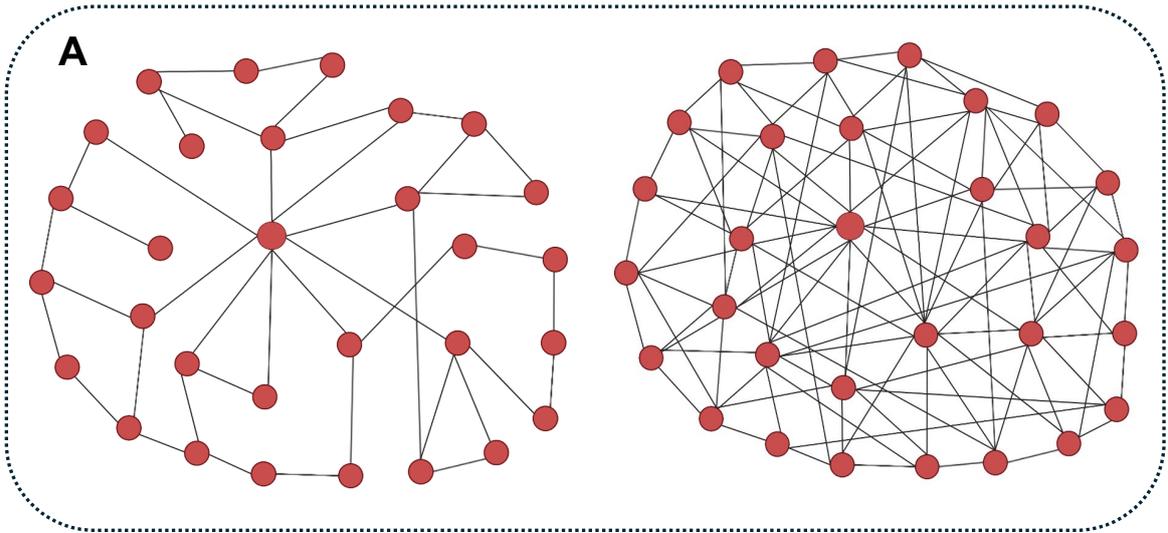

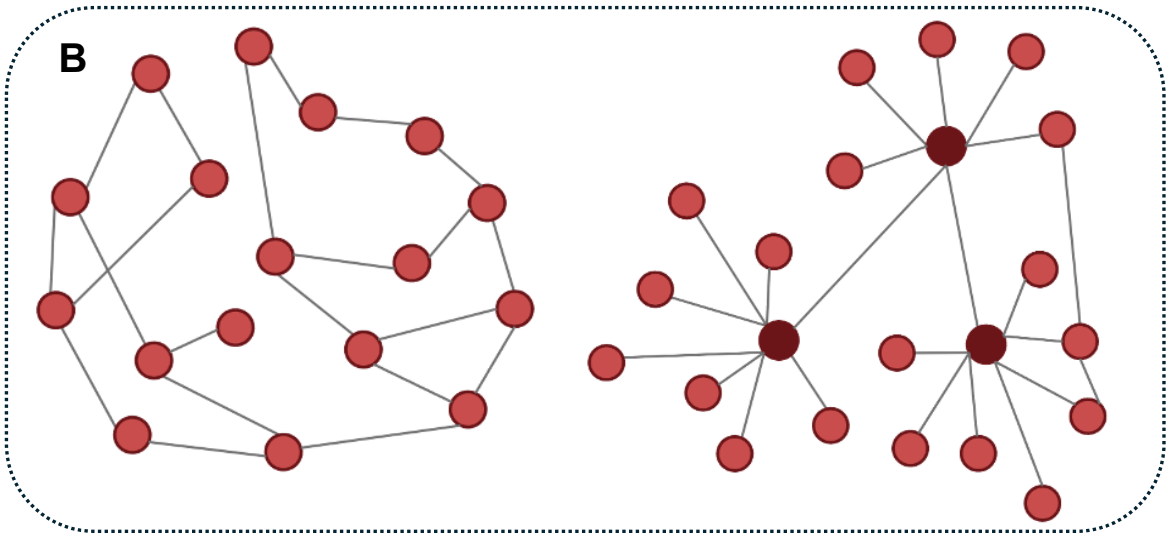

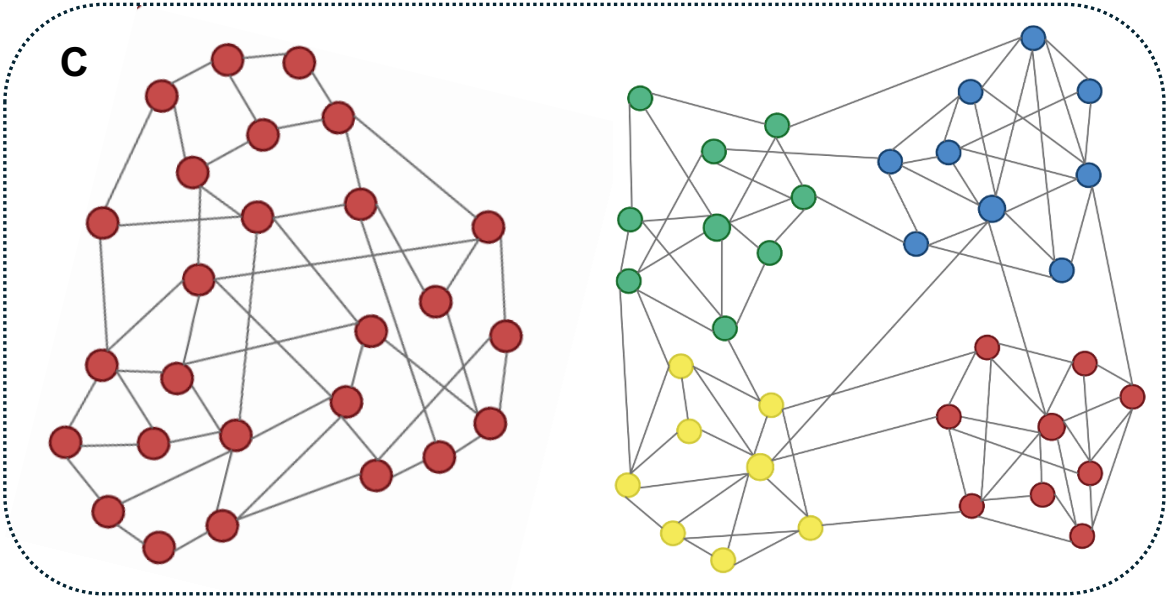

**Figure 5** Structural properties of networks: density, topology, and modularity. **A**: Network density comparison: The left panel depicts a low-density network, while the right panel illustrates a high-density network. **B**: Network topology: The left panel shows a random network, whereas the right panel presents a scale-free network, with darker nodes indicating high-degree hubs. **C**: Modularity differences: The left panel represents a low-modularity network, while the right panel displays a high-modularity network. Node colors correspond to distinct communities, highlighting modular structure.

*Adaptations of macro-scale metrics in TNs*

A core challenge in TN analysis at the macroscale level lies in capturing how global properties evolve over time. Unlike static networks, which assume a fixed topology, TNs require adapted measures that account for temporal dynamics. Various approaches have been developed to track network evolution. One common strategy involves time discretisation, where the network is partitioned into a series of snapshots [11,67]. Standard metrics - such as degree distribution, density or clustering coefficient - are computed independently at each time point, generating temporal trajectories that reveal dynamic shifts in node influence or connectivity. For example, a sharp increase in APL over time may indicate the loss or mutation of a hub gene, disrupting network-wide structure – an effect observed in both metabolic and PPI networks [12,235,236]. Alternatively, a sliding window technique can be used to construct overlapping subnetworks by segmenting temporal data into successive time windows [237,238]. Metrics are then computed within each window, enabling a smoother temporal trend of network evolution. A primary limitation of this approach is the need to predefine the window size, which can bias outcomes [94,239]. While this constraint poses a more significant issue in domains with high-frequency data (e.g. finance or social sciences), its impact is often mitigated in biological networks, where measurements are typically sparser. In both time discretisation and sliding window approaches, traditional macroscale metrics must be conceptually reformulated to respect temporal ordering. For instance, the temporal clustering coefficient considers time-respecting

triads – triplets of nodes where interactions occur in a causally ordered sequence [240,241]. Similarly, temporal shortest paths introduce a key constraint: each consecutive edge in the path must occur after the previous one. This ensures that paths respect causality and chronological progression, yielding more biologically plausible models for signaling or regulatory cascades [36,242].

An alternative modelling strategy involves temporal network aggregation, where a single, cumulative network - also known as a supergraph or union graph - is constructed by assigning edge weights based on their frequency of occurrence over time [11,30,243,244]. While this approach summarises long-term connectivity patterns, it presents limitations [245,246]. Aggregation may obscure the timing of specific events, making it difficult to detect intermittent interactions that could be biologically or structurally significant [247]. Furthermore, it treats all interaction frequencies equally, regardless of context [40,248]. For instance, two node pairs with identical aggregated edge weights may reflect vastly different dynamics—one arising from repeated weak interactions and the other from a few strong, temporally localised events [249]. This ambiguity is especially problematic in biological systems such as PPI networks, where distinguishing between persistent complexes and transient regulatory bindings is crucial. Without careful interpretation, aggregation may overemphasise noisy interactions or underrepresent transient but functionally significant events. The three main strategies discussed are summarised in Figure 6.

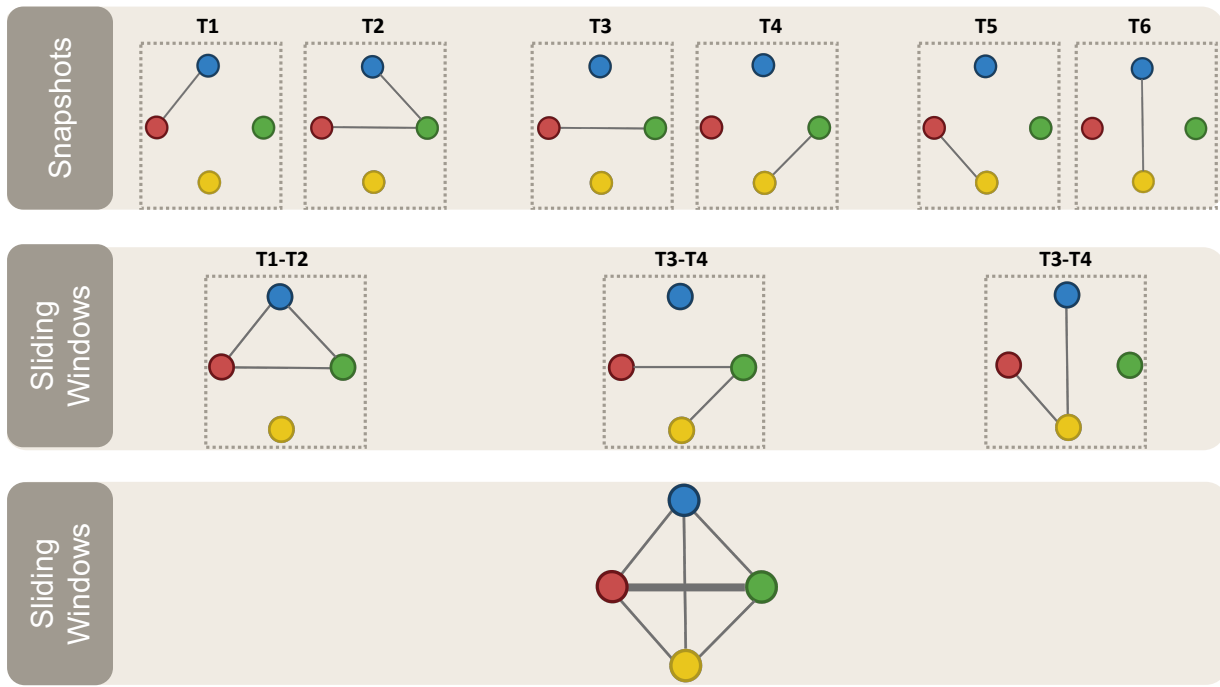

**Figure 6** Temporal network analysis strategies: snapshots, sliding windows, and aggregation. **Top-Panel**: Time-discretised snapshots – The network is partitioned into sequential, non-overlapping time intervals (T1–T6), with each snapshot reflecting the network structure at a specific time point. Standard metrics (e.g., degree, clustering, APL) can be computed per snapshot to track temporal evolution. **Middle-Panel**: Sliding window subnetworks – Overlapping temporal windows (T1–T2, T3–T4, T5–T6) are used to capture local trends while smoothing abrupt fluctuations. Metrics are computed within each window to generate continuous trajectories of change. **Bottom-Panel**: Aggregated network – All temporal interactions are combined into a single cumulative graph, where edge weights represent interaction frequency. While summarising long-term connectivity, this approach may obscure temporally localised or transient events.

## Temporal Network Analysis Approaches

Temporal network analysis is a broad and rapidly evolving field that is inherently complex due to the integration of time-dependent dynamics. Analytical approaches can generally be divided into two categories: direct and indirect. In direct approaches, traditional network metrics are extended to incorporate temporal information—such as temporal centrality and time-respecting shortest paths, both discussed earlier [250,251]. In contrast, indirect approaches transform temporal networks into alternative representations, such as multilayer networks, static snapshots, or proxy structures, which enable the application of established tools like network alignment or graph similarity methods. The choice of approach typically depends on the specific analytical task, with common objectives including the study of structural changes, similarity measures, community evolution, and predictive modeling.

*Structural changes*

A core question in TN analysis is how the topology of networks evolves over time. Researchers investigate local and global structural changes—such as trends in network growth, fragmentation, or rewiring, which involve significant changes in node connections [252,253]. In biological systems, shifts in connectivity may correspond to functional transitions, offering insights into key pathway alterations between physiological states, such as healthy vs disease. For instance, rewiring within cellular molecular networks has been implicated in driving distinct phenotypic outcomes, including embryonic development and cellular differentiation [254]. To explore such dynamics, snapshot-based analysis is often used, where network metrics like degree distribution and clustering coefficient are calculated at each time point and compared across snapshots [237,255]. Other methods include differential network analysis algorithms, which are widely regarded as effective tools for tracking topological changes and identifying recurring patterns [256,257]. In parallel multilayer network models enable the study of both intra-layer edges (within a time point) and inter-layer edges (across time points) [258].

*Comparative analyses*

Comparing and quantifying the similarity between networks at different time points helps assess whether a system remains stable or undergoes significant structural changes. Several methods are available for this purpose, including the commonly employed proximity statistic Jaccard Index (which measures edge overlap), graph edit distance (reflecting the minimal transformation required between two graphs), and graph kernels (functions designed to compare graph structures based on their features) [259,260]. Additionally, network alignment techniques can be applied to compare networks and identify similarities. Two alignment strategies are typically used: local network alignment, which compares the building blocks of two networks, such as motifs; and global network alignment, which seeks to align entire networks by maximising structural overlap (superimposition) in a way that preserves their overall organisation [261,262]. Each of these approaches constitutes a rich research domain in its own right. While a detailed discussion is beyond the scope of this review, several excellent surveys provide comprehensive overviews [263–265].

*Community evolution*

This task focuses on how communities form, evolve, and dissolve within temporal networks. Key phenomena include community persistence, emergence, splitting, merging, and death, as outlined earlier. Several advanced algorithms have been adapted to capture these dynamics. For instance, the Infomap algorithm is a flow-based community detection method that tracks the evolution of communities across time [266]. The Generalised Louvain method, on the other hand, extends modularity-based community detection by optimising both within-layer and between-layer structures [267]. Conversely, Dynamic Stochastic Block Models (DSBMs) provide a probabilistic framework to model community transitions across networks by inferring latent group structures and their evolution over time [268,269]. Additional methods include identifying persistent cliques across time points [270–272], detecting temporal communities via frequent pattern mining (e.g. ABACUS) [273,274], or modeling smooth community transitions through evolutionary clustering, which balances temporal stability with clustering quality (e.g. Facetnet) [275,276].

*Predictive modeling*

Predictive modeling represents a frontier in temporal network analysis. By leveraging historical network data, AI-driven models can forecast future states, such as node influence, structural changes, anomaly detection or information diffusion [277]. A variety of machine-learning techniques have been developed for this purpose. A central strategy involves embedding networks in low-dimensional spaces to capture their structural and temporal properties – a field of study in its own right [278].

Node2vec and DeepWalk are state-of-the-art embedding tools that learn node representations via random walks [279,280]. Conversely, Structural Deep Network Embedding (SDNE) employs deep autoencoders with adjacency matrices and proximity metrics [281]. These algorithms were originally designed for static graphs but have since been extended to temporal settings through dynamic adaptations. For instance, Continuous-Time Dynamic Network Embeddings (CTDNE) learn node-embeddings through time-respecting temporal walks, replacing the time-agnostic random walks used in DeepWalk [282].

Further, deep-learning models like Graph Neural Networks (GNNs) use gradient-based learning to capture nonlinear relationships in graph structures [283]. More recently, Dynamic Graph Neural Networks (DyGNN) have been developed to jointly model temporal and structural aspects and have proven effective in forecasting future interactions in complex systems [284]. While a full discussion of these approaches lies beyond the scope of this review, their emergence illustrates the growing synergy between TN analysis and AI-powered prediction [285,286].

## Conclusion

While static network analysis has provided foundational groundwork for understanding biological systems, temporal network analysis integrates a vital dimension – time – and offers a more nuanced and comprehensive view. At the microscale, it reveals how the roles of individual components shift and fluctuate in response to internal and external cues. At the mesoscale, it captures the dynamics and reconfiguration of evolving functional modules, such as motifs and communities. At the macroscale, it uncovers how large-scale connectivity patterns evolve and information flow, offering insight into system-wide behaviors such as robustness or phase transitions

Across these levels, a unifying insight emerges: biological function is not merely encoded in structure but in how that structure changes over time. Analytical approaches—ranging from similarity metrics and alignment techniques to AI-driven predictive models—demonstrate that integrating temporal data leads to more accurate, context-sensitive interpretations of biological phenomena. Yet, methodological challenges remain, particularly in balancing computational scalability with biological realism.

As the field continues to evolve, the convergence of dynamic network analysis with advancements in AI, high-throughput experimentation, and causal modelling offers an exciting trajectory for future research. Turning these integrated insights into predictive and explanatory tools will require sustained interdisciplinary collaboration, but such efforts are key to deepening our understanding of dynamic biological systems.

## Funding

This research received no external funding.

## Acknowledgments

I wish to record my deep sense of gratitude and profound thanks to my supervisor A/Prof. Fatemeh Vafaee for her guidance and constant encouragement. I also want to thank my colleagues in Vafaee Lab - UNSW Biomedical AI, for their support and insightful comments on this project.

# References

1. Chuang H-Y, Hofree M, Ideker T. A decade of systems biology. Annu. Rev. Cell Dev. Biol. 2010; 26:721–744

2. Barabási A-L, Albert R, Jeong H. Scale-free characteristics of random networks: the topology of the world-wide web. Phys. Stat. Mech. Its Appl. 2000; 281:69–77

3. Kulandaisamy A, Lathi V, ViswaPoorani K, et al. Important amino acid residues involved in folding and binding of protein–protein complexes. Int. J. Biol. Macromol. 2017; 94:438–444

4. Nooren IM, Thornton JM. Diversity of protein–protein interactions. EMBO J. 2003; 22:3486–3492

5. Milroy L, Llamas C. Social networks. Handb. Lang. Var. Change 2013; 407–427

6. Kempe D, Kleinberg J, Tardos É. Maximizing the spread of influence through a social network. 2003; 137–146

7. Tabassum S, Pereira FS, Fernandes S, et al. Social network analysis: An overview. Wiley Interdiscip. Rev. Data Min. Knowl. Discov. 2018; 8:e1256

8. Börner K, Sanyal S, Vespignani A. Network science. Annu Rev Inf Sci Technol 2007; 41:537–607

9. Boccaletti S, Latora V, Moreno Y, et al. Complex networks: Structure and dynamics. Phys. Rep. 2006; 424:175–308

10. Newman M. Networks: An introduction. 2010: Oxford university press. Artif Life 2012; 18:241–242

11. Holme P, Saramäki J. Temporal networks. Phys. Rep. 2012; 519:97–125

12. Albert R, Barabási A-L. Statistical mechanics of complex networks. Rev. Mod. Phys. 2002; 74:47

13. Strogatz SH. Exploring complex networks. nature 2001; 410:268–276

14. Pastor-Satorras R, Castellano C, Van Mieghem P, et al. Epidemic processes in complex networks. Rev. Mod. Phys. 2015; 87:925–979

15. Cho S-Y, Park S-G, Lee D-H, et al. Protein-protein interaction networks: from interactions to networks. BMB Rep. 2004; 37:45–52

16. Safari-Alighiarloo N, Taghizadeh M, Rezaei-Tavirani M, et al. Protein-protein interaction networks (PPI) and complex diseases. Gastroenterol. Hepatol. Bed Bench 2014; 7:17

17. Vella D, Marini S, Vitali F, et al. MTGO: PPI network analysis via topological and functional module identification. Sci. Rep. 2018; 8:5499

18. Koh GC, Porras P, Aranda B, et al. Analyzing protein–protein interaction networks. J. Proteome Res. 2012; 11:2014–2031

19. Lazer D, Pentland A, Adamic L, et al. Computational social science. Science 2009; 323:721–723

20. Bray D. Molecular networks: the top-down view. Science 2003; 301:1864–1865

21. Alm E, Arkin AP. Biological networks. Curr. Opin. Struct. Biol. 2003; 13:193–202

22. Liu C, Ma Y, Zhao J, et al. Computational network biology: data, models, and applications. Phys. Rep. 2020; 846:1–66

23. Bornholdt S, Schuster HG. Handbook of graphs and networks. 2001;

24. Newman M. Networks. 2018;

25. Newman ME. The structure of scientific collaboration networks. Proc. Natl. Acad. Sci. 2001; 98:404–409

26. Newman ME. The mathematics of networks. New Palgrave Encycl. Econ. 2008; 2:1–12

27. Newman ME. Analysis of weighted networks. Phys. Rev. E—Statistical Nonlinear Soft Matter Phys. 2004; 70:056131

28. Casteigts A, Flocchini P, Quattrociocchi W, et al. Time-varying graphs and dynamic networks. Int. J. Parallel Emergent Distrib. Syst. 2012; 27:387–408

29. Lentz HH, Selhorst T, Sokolov IM. Unfolding accessibility provides a macroscopic approach to temporal networks. Phys. Rev. Lett. 2013; 110:118701

30. Masuda N, Lambiotte R. A guide to temporal networks. 2016;

31. Ghosh D, Frasca M, Rizzo A, et al. The synchronized dynamics of time-varying networks. Phys. Rep. 2022; 949:1–63

32. Tyson JJ, Laomettachit T, Kraikivski P. Modeling the dynamic behavior of biochemical regulatory networks. J. Theor. Biol. 2019; 462:514–527

33. Sekara V, Stopczynski A, Lehmann S. Fundamental structures of dynamic social networks. Proc. Natl. Acad. Sci. 2016; 113:9977–9982

34. Takaffoli M, Fagnan J, Sangi F, et al. Tracking changes in dynamic information networks. 2011; 94–101

35. Barrat A, Barthelemy M, Vespignani A. Dynamical processes on complex networks. 2008;

36. Wu H, Cheng J, Huang S, et al. Path problems in temporal graphs. Proc. VLDB Endow. 2014; 7:721–732

37. Barros CD, Mendonça MR, Vieira AB, et al. A survey on embedding dynamic graphs. ACM Comput. Surv. CSUR 2021; 55:1–37

38. Powell WW, White DR, Koput KW, et al. Network dynamics and field evolution: The growth of interorganizational collaboration in the life sciences. Am. J. Sociol. 2005; 110:1132–1205

39. Kossinets G, Watts DJ. Empirical analysis of an evolving social network. science 2006; 311:88–90

40. Masuda N, Klemm K, Eguíluz VM. Temporal networks: slowing down diffusion by long lasting interactions. Phys. Rev. Lett. 2013; 111:188701

41. Eisenstein M. Big data: The power of petabytes. Nature 2015; 527:S2–S4

42. Greenfield A, Madar A, Ostrer H, et al. DREAM4: Combining genetic and dynamic information to identify biological networks and dynamical models. PloS One 2010; 5:e13397

43. Palsson BØ. Systems biology: simulation of dynamic network states. 2011;

44. Jin Z, El-Deiry WS. Overview of cell death signaling pathways. Cancer Biol. Ther. 2005; 4:147–171

45. Kawasaki T, Kawai T. Toll-like receptor signaling pathways. Front. Immunol. 2014; 5:461

46. Granot I, Dekel N. Cell-to-cell communication in the ovarian follicle: developmental and hormonal regulation of the expression of connexin43. Hum. Reprod. 1998; 13:85–97

47. Lovinger DM. Communication networks in the brain: neurons, receptors, neurotransmitters, and alcohol. Alcohol Res. Health 2008; 31:196

48. Karlebach G, Shamir R. Modelling and analysis of gene regulatory networks. Nat. Rev. Mol. Cell Biol. 2008; 9:770–780

49. Luscombe NM, Madan Babu M, Yu H, et al. Genomic analysis of regulatory network dynamics reveals large topological changes. Nature 2004; 431:308–312

50. Hommes D, Peppelenbosch M, Van Deventer S. Mitogen activated protein (MAP) kinase signal transduction pathways and novel anti-inflammatory targets. Gut 2003; 52:144–151

51. Zhang W, Liu HT. MAPK signal pathways in the regulation of cell proliferation in mammalian cells. Cell Res. 2002; 12:9–18

52. Zheng C-F, Guan K-L. Activation of MEK family kinases requires phosphorylation of two conserved Ser/Thr residues. EMBO J. 1994; 13:1123–1131

53. Sanchez JN, Wang T, Cohen MS. BRAF and MEK inhibitors: use and resistance in BRAF-mutated cancers. Drugs 2018; 78:549–566

54. Sun Q, Wang W. Structures of BRAF–MEK1–14-3-3 sheds light on drug discovery. Signal Transduct. Target. Ther. 2019; 4:59

55. Neuzillet C, Tijeras-Raballand A, de Mestier L, et al. MEK in cancer and cancer therapy. Pharmacol. Ther. 2014; 141:160–171

56. Sahni N, Yi S, Taipale M, et al. Widespread macromolecular interaction perturbations in human genetic disorders. Cell 2015; 161:647–660

57. Zhong Q, Simonis N, Li Q, et al. Edgetic perturbation models of human inherited disorders. Mol. Syst. Biol. 2009; 5:321

58. Barabási A-L, Gulbahce N, Loscalzo J. Network medicine: a network-based approach to human disease. Nat. Rev. Genet. 2011; 12:56–68

59. Intlekofer AM, Takemoto N, Wherry EJ, et al. Effector and memory CD8+ T cell fate coupled by T-bet and eomesodermin. Nat. Immunol. 2005; 6:1236–1244

60. Rudensky AY. Regulatory T cells and Foxp3. Immunol. Rev. 2011; 241:260–268

61. Shi H, Yan K-K, Ding L, et al. Network approaches for dissecting the immune system. Iscience 2020; 23:

62. Albert R, Barabási A-L. Topology of evolving networks: local events and universality. Phys. Rev. Lett. 2000; 85:5234

63. Yook S-H, Jeong H, Barabási A-L, et al. Weighted evolving networks. Phys. Rev. Lett. 2001; 86:5835

64. Grindrod P, Higham DJ. Evolving graphs: dynamical models, inverse problems and propagation. Proc. R. Soc. Math. Phys. Eng. Sci. 2010; 466:753–770

65. Rossi RA, Gallagher B, Neville J, et al. Modeling dynamic behavior in large evolving graphs. 2013; 667–676

66. Holme P. Modern temporal network theory: a colloquium. Eur. Phys. J. B 2015; 88:1–30

67. Li A, Cornelius SP, Liu Y-Y, et al. The fundamental advantages of temporal networks. Science 2017; 358:1042–1046

68. Kostakos V. Temporal graphs. Phys. Stat. Mech. Its Appl. 2009; 388:1007–1023

69. Michail O. An introduction to temporal graphs: An algorithmic perspective. Internet Math. 2016; 12:239–280

70. Šiljak D. Dynamic graphs. Nonlinear Anal. Hybrid Syst. 2008; 2:544–567

71. Trivedi R, Farajtabar M, Biswal P, et al. Dyrep: Learning representations over dynamic graphs. 2019;

72. Zadeh LA. Time-varying networks, I. Proc. IRE 1961; 49:1488–1503

73. Kolar M, Song L, Ahmed A, et al. Estimating time-varying networks. Ann. Appl. Stat. 2010; 94–123

74. Santoro N, Quattrociocchi W, Flocchini P, et al. Time-varying graphs and social network analysis: Temporal indicators and metrics. ArXiv Prepr. ArXiv11020629 2011;

75. Nicosia V, Tang J, Musolesi M, et al. Components in time-varying graphs. Chaos Interdiscip. J. Nonlinear Sci. 2012; 22:

76. Mucha PJ, Porter MA. Communities in multislice voting networks. Chaos Interdiscip. J. Nonlinear Sci. 2010; 20:

77. Carchiolo V, Longheu A, Malgeri M, et al. Communities unfolding in multislice networks. 2011; 187–195

78. Farkas M. Dynamical models in biology. 2001;

79. Prill RJ, Marbach D, Saez-Rodriguez J, et al. Towards a rigorous assessment of systems biology models: the DREAM3 challenges. PloS One 2010; 5:e9202

80. Aggarwal C, Subbian K. Evolutionary network analysis: A survey. ACM Comput. Surv. CSUR 2014; 47:1–36

81. Yu W, Aggarwal CC, Wang W. Temporally factorized network modeling for evolutionary network analysis. 2017; 455–464

82. Sarkar P, Moore AW. Dynamic social network analysis using latent space models. Acm Sigkdd Explor. Newsl. 2005; 7:31–40

83. Proskurnikov AV, Tempo R. A tutorial on modeling and analysis of dynamic social networks. Part I. Annu. Rev. Control 2017; 43:65–79

84. Vogels TP, Rajan K, Abbott LF. Neural network dynamics. Annu Rev Neurosci 2005; 28:357–376

85. Trung HT, Toan NT, Van Vinh T, et al. A comparative study on network alignment techniques. Expert Syst. Appl. 2020; 140:112883

86. Li Y, Gu C, Dullien T, et al. Graph Matching Networks for Learning the Similarity of Graph Structured Objects. Proc. 36th Int. Conf. Mach. Learn. 2019; 3835–3845

87. Ling X, Wu L, Wang S, et al. Multilevel Graph Matching Networks for Deep Graph Similarity Learning. IEEE Trans. Neural Netw. Learn. Syst. 2023; 34:799–813

88. Kollias G, Mohammadi S, Grama A. Network similarity decomposition (nsd): A fast and scalable approach to network alignment. IEEE Trans. Knowl. Data Eng. 2011; 24:2232–2243

89. Watts DJ, Strogatz SH. Collective dynamics of 'small-world'networks. nature 1998; 393:440–442

90. Kinsley AC, Rossi G, Silk MJ, et al. Multilayer and multiplex networks: An introduction to their use in veterinary epidemiology. Front. Vet. Sci. 2020; 7:596

91. Boccaletti S, Bianconi G, Criado R, et al. The structure and dynamics of multilayer networks. Phys. Rep. 2014; 544:1–122

92. Zhang H. Temporal Subgraph Matching Method for Multi-Connected Temporal Graph. Inf. Sci. 2024; 121320

93. Kempe D, Kleinberg J, Kumar A. Connectivity and inference problems for temporal networks. 2000; 504–513

94. Sikdar S, Ganguly N, Mukherjee A. Time series analysis of temporal networks. Eur. Phys. J. B 2016; 89:1–11

95. Opsahl T, Agneessens F, Skvoretz J. Node centrality in weighted networks: Generalizing degree and shortest paths. Soc. Netw. 2010; 32:245–251

96. Kim H, Anderson R. Temporal node centrality in complex networks. Phys. Rev. E 2012; 85:026107

97. Zhong S, Zhang H, Deng Y. Identification of influential nodes in complex networks: A local degree dimension approach. Inf. Sci. 2022; 610:994–1009

98. Yu E-Y, Fu Y, Chen X, et al. Identifying critical nodes in temporal networks by network embedding. Sci. Rep. 2020; 10:12494

99. Freeman LC. Centrality in social networks: Conceptual clarification. Soc. Netw. Crit. Concepts Sociol. Lond. Routledge 2002; 1:238–263

100. Iyer S, Killingback T, Sundaram B, et al. Attack robustness and centrality of complex networks. PloS One 2013; 8:e59613

101. Jeong H, Mason SP, Barabási A-L, et al. Lethality and centrality in protein networks. Nature 2001; 411:41–42

102. He X, Zhang J. Why do hubs tend to be essential in protein networks? PLoS Genet. 2006; 2:e88

103. Freeman LC. A set of measures of centrality based on betweenness. Sociometry 1977; 35–41

104. Gallo G, Pallottino S. Shortest path methods: A unifying approach. Netflow Pisa 1986; 38–64

105. Peyré G, Péchaud M, Keriven R, et al. Geodesic methods in computer vision and graphics. Found. Trends® Comput. Graph. Vis. 2010; 5:197–397

106. Bouttier J, Di Francesco P, Guitter E. Geodesic distance in planar graphs. Nucl. Phys. B 2003; 663:535–567

107. Hamed M, Spaniol C, Zapp A, et al. Integrative network-based approach identifies key genetic elements in breast invasive carcinoma. BMC Genomics 2015; 16:1–14

108. Nazarieh M, Helms V. TopControl: A tool to prioritize candidate disease-associated genes based on topological network features. Sci. Rep. 2019; 9:19472

109. Nazarieh M, Wiese A, Will T, et al. Identification of key player genes in gene regulatory networks. BMC Syst. Biol. 2016; 10:1–12

110. Lavarenne J, Guyomarc'h S, Sallaud C, et al. The spring of systems biology-driven breeding. Trends Plant Sci. 2018; 23:706–720

111. Nacher JC, Akutsu T. Minimum dominating set-based methods for analyzing biological networks. Methods 2016; 102:57–63

112. Tsalouchidou I, Baeza-Yates R, Bonchi F, et al. Temporal betweenness centrality in dynamic graphs. Int. J. Data Sci. Anal. 2020; 9:257–272

113. Zaoli S, Mazzarisi P, Lillo F. Betweenness centrality for temporal multiplexes. Sci. Rep. 2021; 11:4919

114. Machens A, Gesualdo F, Rizzo C, et al. An infectious disease model on empirical networks of human contact: bridging the gap between dynamic network data and contact matrices. BMC Infect. Dis. 2013; 13:1–15

115. Husein I, Mawengkang H, Suwilo S. Modeling the transmission of infectious disease in a dynamic network. 2019; 1255:012052

116. Xie J, Yang F, Wang J, et al. DNF: a differential network flow method to identify rewiring drivers for gene regulatory networks. Neurocomputing 2020; 410:202–210

117. Pan Z, Li L, Fang Q, et al. Analysis of dynamic molecular networks for pancreatic ductal adenocarcinoma progression. Cancer Cell Int. 2018; 18:1–18

118. Hugues S, Fetler L, Bonifaz L, et al. Distinct T cell dynamics in lymph nodes during the induction of tolerance and immunity. Nat. Immunol. 2004; 5:1235–1242

119. Jaeger M, Anastasio A, Chamy L, et al. Light-inducible T cell engagers trigger, tune, and shape the activation of primary T cells. Proc. Natl. Acad. Sci. 2023; 120:e2302500120

120. Erwin DH, Davidson EH. The evolution of hierarchical gene regulatory networks. Nat. Rev. Genet. 2009; 10:141–148

121. Stadhouders R, Vidal E, Serra F, et al. Transcription factors orchestrate dynamic interplay between genome topology and gene regulation during cell reprogramming. Nat. Genet. 2018; 50:238–249

122. Heinz S, Romanoski CE, Benner C, et al. The selection and function of cell type-specific enhancers. Nat. Rev. Mol. Cell Biol. 2015; 16:144–154

123. Schoenrock A, Burnside D, Moteshareie H, et al. Evolution of protein-protein interaction networks in yeast. PLoS One 2017; 12:e0171920

124. Lu X, Jain VV, Finn PW, et al. Hubs in biological interaction networks exhibit low changes in expression in experimental asthma. Mol. Syst. Biol. 2007; 3:98

125. Tummino PJ, Copeland RA. Residence time of receptor– ligand complexes and its effect on biological function. Biochemistry 2008; 47:5481–5492

126. Kusumi A, Tsunoyama TA, Suzuki KG, et al. Transient, nano-scale, liquid-like molecular assemblies coming of age. Curr. Opin. Cell Biol. 2024; 89:102394

127. Bagatell R, Whitesell L. Altered Hsp90 function in cancer: a unique therapeutic opportunity. Mol. Cancer Ther. 2004; 3:1021–1030

128. Wang W, Sreekumar PG, Valluripalli V, et al. Protein polymer nanoparticles engineered as chaperones protect against apoptosis in human retinal pigment epithelial cells. J. Controlled Release 2014; 191:4–14

129. Masquelier T, Thorpe SJ. Unsupervised learning of visual features through spike timing dependent plasticity. PLoS Comput. Biol. 2007; 3:e31

130. Zeng G, Huang X, Jiang T, et al. Short-term synaptic plasticity expands the operational range of long-term synaptic changes in neural networks. Neural Netw. 2019; 118:140–147

131. Kashyap G, Bapat D, Das D, et al. Synapse loss and progress of Alzheimer's disease-A network model. Sci. Rep. 2019; 9:6555

132. Carter SL, Brechbühler CM, Griffin M, et al. Gene co-expression network topology provides a framework for molecular characterization of cellular state. Bioinformatics 2004; 20:2242–2250

133. Chaplin DD. Overview of the immune response. J. Allergy Clin. Immunol. 2010; 125:S3–S23

134. Zhang T, Gao Y, Qiu L, et al. Distributed time-respecting flow graph pattern matching on temporal graphs. World Wide Web 2020; 23:609–630

135. Van Der Wijst MG, de Vries DH, Brugge H, et al. An integrative approach for building personalized gene regulatory networks for precision medicine. Genome Med. 2018; 10:1–15

136. Chan SS-K, Kyba M. What is a master regulator? J. Stem Cell Res. Ther. 2013; 3:

137. Itzkovitz S, Milo R, Kashtan N, et al. Subgraphs in random networks. Phys. Rev. E 2003; 68:026127

138. Ribeiro P, Paredes P, Silva ME, et al. A survey on subgraph counting: concepts, algorithms, and applications to network motifs and graphlets. ACM Comput. Surv. CSUR 2021; 54:1–36

139. Bondy JA, Murty USR. Graph theory. 2008;

140. Estrada E, Rodriguez-Velazquez JA. Subgraph centrality in complex networks. Phys. Rev. E—Statistical Nonlinear Soft Matter Phys. 2005; 71:056103

141. Estrada E, Rodríguez-Velázquez JA. Subgraph centrality and clustering in complex hyper-networks. Phys. Stat. Mech. Its Appl. 2006; 364:581–594

142. Pržulj N, Corneil DG, Jurisica I. Modeling interactome: scale-free or geometric? Bioinformatics 2004; 20:3508–3515

143. Yaveroglu ON, Fitzhugh SM, Kurant M, et al. ergm.graphlets: a package for ERG modeling based on graphlet statistics. ArXiv Prepr. ArXiv14057348 2014;

144. Milenković T, Pržulj N. Uncovering biological network function via graphlet degree signatures. Cancer Inform. 2008; 6:CIN-S680

145. Trpevski I, Dimitrova T, Boshkovski T, et al. Graphlet characteristics in directed networks. Sci. Rep. 2016; 6:37057

146. Milo R, Itzkovitz S, Kashtan N, et al. Superfamilies of evolved and designed networks. Science 2004; 303:1538–1542

147. Alon U. An introduction to systems biology: design principles of biological circuits. 2019;

148. Milo R, Shen-Orr S, Itzkovitz S, et al. Network motifs: simple building blocks of complex networks. Science 2002; 298:824–827

149. Li Y, Lee KK, Walsh S, et al. Establishing glucose-and ABA-regulated transcription networks in Arabidopsis by microarray analysis and promoter classification using a Relevance Vector Machine. Genome Res. 2006; 16:414–427

150. Lee TI, Rinaldi NJ, Robert F, et al. Transcriptional regulatory networks in Saccharomyces cerevisiae. science 2002; 298:799–804

151. Saddic LA, Huvermann B, Bezhani S, et al. The LEAFY target LMI1 is a meristem identity regulator and acts together with LEAFY to regulate expression of CAULIFLOWER. 2006;

152. Odom DT, Zizlsperger N, Gordon DB, et al. Control of pancreas and liver gene expression by HNF transcription factors. Science 2004; 303:1378–1381

153. Shen-Orr SS, Milo R, Mangan S, et al. Network motifs in the transcriptional regulation network of Escherichia coli. Nat. Genet. 2002; 31:64–68

154. Mangan S, Alon U. Structure and function of the feed-forward loop network motif. Proc. Natl. Acad. Sci. 2003; 100:11980–11985

155. McAdams HH, Shapiro L. Circuit simulation of genetic networks. Science 1995; 269:650–656

156. Ma'ayan A, Jenkins SL, Neves S, et al. Formation of regulatory patterns during signal propagation in a mammalian cellular network. Science 2005; 309:1078–1083

157. Apte AA, Cain JW, Bonchev DG, et al. Cellular automata simulation of topological effects on the dynamics of feed-forward motifs. J. Biol. Eng. 2008; 2:1–12

158. Kashtan N, Itzkovitz S, Milo R, et al. Topological generalizations of network motifs. Phys. Rev. E—Statistical Nonlinear Soft Matter Phys. 2004; 70:031909

159. Piraveenan M, Wimalawarne K, Kasthurirathn D. Centrality and composition of four-node motifs in metabolic networks. Procedia Comput. Sci. 2013; 18:409–418

160. Kepes F. Biological networks. 2007; 3:

161. Yu S, Xu J, Zhang C, et al. Motifs in big networks: Methods and applications. IEEE Access 2019; 7:183322–183338

162. Provan KG, Sebastian JG. Networks within networks: Service link overlap, organizational cliques, and network effectiveness. Acad. Manage. J. 1998; 41:453–463

163. Fadigas I de S, Pereira HB de B. A network approach based on cliques. Phys. Stat. Mech. Its Appl. 2013; 392:2576–2587

164. Luce RD, Perry AD. A method of matrix analysis of group structure. Psychometrika 1949; 14:95–116

165. Baldwin NE, Chesler EJ, Kirov S, et al. Computational, integrative, and comparative methods for the elucidation of genetic coexpression networks. BioMed Res. Int. 2005; 2005:172–180

166. Schmidt MC, Samatova NF, Thomas K, et al. A scalable, parallel algorithm for maximal clique enumeration. J. Parallel Distrib. Comput. 2009; 69:417–428

167. Ouyang Q, Kaplan PD, Liu S, et al. DNA solution of the maximal clique problem. Science 1997; 278:446–449

168. Cazals F, Karande C. A note on the problem of reporting maximal cliques. Theor. Comput. Sci. 2008; 407:564–568

169. Cheng J, Ke Y, Fu AW-C, et al. Finding maximal cliques in massive networks. ACM Trans. Database Syst. TODS 2011; 36:1–34

170. Cheng J, Zhu L, Ke Y, et al. Fast algorithms for maximal clique enumeration with limited memory. 2012; 1240–1248

171. Abu-Khzam FN, Baldwin NE, Langston MA, et al. On the relative efficiency of maximal clique enumeration algorithms, with applications to high-throughput computational biology. 2005; 1–10

172. Wang J, Liu B, Li M, et al. Identifying protein complexes from interaction networks based on clique percolation and distance restriction. BMC Genomics 2010; 11:1–14

173. Arenas A, Fernandez A, Fortunato S, et al. Motif-based communities in complex networks. J. Phys. Math. Theor. 2008; 41:224001

174. Fortunato S. Community detection in graphs. Phys. Rep. 2010; 486:75–174

175. Radicchi F, Castellano C, Cecconi F, et al. Defining and identifying communities in networks. Proc. Natl. Acad. Sci. 2004; 101:2658–2663

176. Porter MA, Onnela J-P, Mucha PJ. Communities in networks. 2009;

177. Coscia M, Giannotti F, Pedreschi D. A classification for community discovery methods in complex networks. Stat. Anal. Data Min. ASA Data Sci. J. 2011; 4:512–546

178. Li X, Wu M, Kwoh C-K, et al. Computational approaches for detecting protein complexes from protein interaction networks: a survey. BMC Genomics 2010; 11:1–19

179. Ni C-C, Lin Y-Y, Luo F, et al. Community detection on networks with Ricci flow. Sci. Rep. 2019; 9:9984

180. Newman ME, Girvan M. Finding and evaluating community structure in networks. Phys. Rev. E 2004; 69:026113

181. Newman ME. Modularity and community structure in networks. Proc. Natl. Acad. Sci. 2006; 103:8577–8582

182. Alcalá-Corona SA, Sandoval-Motta S, Espinal-Enriquez J, et al. Modularity in biological networks. Front. Genet. 2021; 12:701331

183. Valentini G, Paccanaro A, Caniza H, et al. An extensive analysis of disease-gene associations using network integration and fast kernel-based gene prioritization methods. Artif. Intell. Med. 2014; 61:63–78

184. Manipur I, Giordano M, Piccirillo M, et al. Community detection in protein-protein interaction networks and applications. IEEE/ACM Trans. Comput. Biol. Bioinform. 2021; 20:217–237

185. Sevimoglu T, Arga KY. The role of protein interaction networks in systems biomedicine. Comput. Struct. Biotechnol. J. 2014; 11:22–27

186. Peixoto TP, Rosvall M. Modelling sequences and temporal networks with dynamic community structures. Nat. Commun. 2017; 8:582

187. Masuda N, Holme P. Detecting sequences of system states in temporal networks. Sci. Rep. 2019; 9:795

188. Hosseinzadeh MM, Cannataro M, Guzzi PH, et al. Temporal networks in biology and medicine: a survey on models, algorithms, and tools. Netw. Model. Anal. Health Inform. Bioinforma. 2022; 12:10

189. Dell'Amico M, Filippone M, Michiardi P, et al. On user availability prediction and network applications. IEEEACM Trans. Netw. 2014; 23:1300–1313

190. Lucas M, Morris A, Townsend-Teague A, et al. Inferring cell cycle phases from a partially temporal network of protein interactions. Cell Rep. Methods 2023; 3:

191. Kovanen L, Karsai M, Kaski K, et al. Temporal motifs in time-dependent networks. J. Stat. Mech. Theory Exp. 2011; 2011:P11005

192. Liu P, Guarrasi V, Sarıyüce AE. Temporal network motifs: Models, limitations, evaluation. IEEE Trans. Knowl. Data Eng. 2021; 35:945–957

193. Shen X, Li Y, Jiang X, et al. Detecting temporal protein complexes based on neighbor closeness and time course protein interaction networks. 2015; 109–112

194. Zhang Y, Lin H, Yang Z, et al. Construction of dynamic probabilistic protein interaction networks for protein complex identification. BMC Bioinformatics 2016; 17:1–13

195. de Lichtenberg U, Jensen LJ, Brunak S, et al. Dynamic complex formation during the yeast cell cycle. science 2005; 307:724–727

196. Pang K, Sheng H, Ma X. Understanding gene essentiality by finely characterizing hubs in the yeast protein interaction network. Biochem. Biophys. Res. Commun. 2010; 401:112–116

197. Zhang A. Protein interaction networks: computational analysis. 2009;

198. Lin C-C, Hsiang J-T, Wu C-Y, et al. Dynamic functional modules in co-expressed protein interaction networks of dilated cardiomyopathy. BMC Syst. Biol. 2010; 4:1–14

199. Kuhn F, Oshman R. Dynamic networks: models and algorithms. ACM SIGACT News 2011; 42:82–96

200. Kim B, Lee KH, Xue L, et al. A review of dynamic network models with latent variables. Stat. Surv. 2018; 12:105

201. Li C, Maini PK. An evolving network model with community structure. J. Phys. Math. Gen. 2005; 38:9741

202. Chou T, D'Orsogna MR. First passage problems in biology. First-Passage Phenom. Their Appl. 2014; 306–345

203. Singer P, Helic D, Taraghi B, et al. Detecting memory and structure in human navigation patterns using Markov chain models of varying order. PloS One 2014; 9:e102070

204. Palazzi Nieves MJ, Borge-Holthoefer J, Tessone CJ, et al. Macro-and mesoscale pattern interdependencies in complex networks. J. R. Soc. Interface 2019 16 159 2019;

205. Bedru HD, Yu S, Xiao X, et al. Big networks: A survey. Comput. Sci. Rev. 2020; 37:100247

206. Intanagonwiwat C, Estrin D, Govindan R, et al. Impact of network density on data aggregation in wireless sensor networks. 2002; 457–458

207. Bhattacharya S, Sinha S, Dey P, et al. Online social-network sensing models. Comput. Intell. Appl. Text Sentim. Data Anal. 2023; 113–140

208. Jiang P, Singh M. SPICi: a fast clustering algorithm for large biological networks. Bioinformatics 2010; 26:1105–1111

209. Holman J. Dense graphlet statistics of protein interaction and random networks. Biocomput. 2009 2009; 178–189

210. Golbeck J. Chapter 3 - Network Structure and Measures. Anal. Soc. Web 2013; 25–44

211. Barabási A-L, Bonabeau E. Scale-free networks. Sci. Am. 2003; 288:60–69

212. Barabási A-L. Scale-free networks: a decade and beyond. science 2009; 325:412–413

213. Barabasi A-L, Oltvai ZN. Network biology: understanding the cell's functional organization. Nat. Rev. Genet. 2004; 5:101–113

214. Adamic LA, Huberman BA. Power-law distribution of the world wide web. science 2000; 287:2115–2115

215. Zhao J, Xu K. Enhancing the robustness of scale-free networks. J. Phys. Math. Theor. 2009; 42:195003

216. Cohen R, Erez K, Ben-Avraham D, et al. Resilience of the internet to random breakdowns. Phys. Rev. Lett. 2000; 85:4626

217. Albert R, Jeong H, Barabási A-L. Error and attack tolerance of complex networks. nature 2000; 406:378–382

218. Xiao S, Xiao G, Cheng T, et al. Robustness of scale-free networks under rewiring operations. Europhys. Lett. 2010; 89:38002

219. Cooper TF, Morby AP, Gunn A, et al. Effect of random and hub gene disruptions on environmental and mutational robustness in Escherichia coli. Bmc Genomics 2006; 7:1–11

220. Li Y, Shang Y, Yang Y. Clustering coefficients of large networks. Inf. Sci. 2017; 382:350–358

221. Soffer SN, Vazquez A. Network clustering coefficient without degree-correlation biases. Phys. Rev. E—Statistical Nonlinear Soft Matter Phys. 2005; 71:057101

222. Kong X, Shi Y, Yu S, et al. Academic social networks: Modeling, analysis, mining and applications. J. Netw. Comput. Appl. 2019; 132:86–103

223. Gilarranz LJ, Rayfield B, Liñán-Cembrano G, et al. Effects of network modularity on the spread of perturbation impact in experimental metapopulations. Science 2017; 357:199–201

224. Ravasz E, Somera AL, Mongru DA, et al. Hierarchical organization of modularity in metabolic networks. science 2002; 297:1551–1555

225. Lovejoy WS, Loch CH. Minimal and maximal characteristic path lengths in connected sociomatrices. Soc. Netw. 2003; 25:333–347

226. Embar V, Handen A, Ganapathiraju MK. Is the average shortest path length of gene set a reflection of their biological relatedness? J. Bioinform. Comput. Biol. 2016; 14:1660002

227. Doncheva NT, Kacprowski T, Albrecht M. Recent approaches to the prioritization of candidate disease genes. Wiley Interdiscip. Rev. Syst. Biol. Med. 2012; 4:429–442

228. Radivojac P, Peng K, Clark WT, et al. An integrated approach to inferring gene–disease associations in humans. Proteins Struct. Funct. Bioinforma. 2008; 72:1030–1037

229. Zhang L, Li X, Tai J, et al. Predicting candidate genes based on combined network topological features: a case study in coronary artery disease. PloS One 2012; 7:e39542

230. Köhler S, Bauer S, Horn D, et al. Walking the interactome for prioritization of candidate disease genes. Am. J. Hum. Genet. 2008; 82:949–958

231. Masuda N, Sakaki M, Ezaki T, et al. Clustering coefficients for correlation networks. Front. Neuroinformatics 2018; 12:7

232. Rubinov M, Sporns O. Complex network measures of brain connectivity: uses and interpretations. Neuroimage 2010; 52:1059–1069

233. Newman ME. The structure and function of complex networks. SIAM Rev. 2003; 45:167–256

234. Foster JG, Foster DV, Grassberger P, et al. Edge direction and the structure of networks. Proc. Natl. Acad. Sci. 2010; 107:10815–10820

235. Vogelstein B, Lane D, Levine AJ. Surfing the p53 network. Nature 2000; 408:307–310

236. Jeong H, Tombor B, Albert R, et al. The large-scale organization of metabolic networks. Nature 2000; 407:651–654

237. Moctar AOM, Sarr[1] I, Tanzouak JV. Snapshot setting for Temporal Networks Analysis. 2019; 275:98

238. Hulovatyy Y, Chen H, Milenković T. Exploring the structure and function of temporal networks with dynamic graphlets. Bioinformatics 2015; 31:i171–i180

239. Jordan DG, Winer ES, Salem T. The current status of temporal network analysis for clinical science: Considerations as the paradigm shifts? J. Clin. Psychol. 2020; 76:1591–1612

240. Cui J, Zhang Y-Q, Li X. On the clustering coefficients of temporal networks and epidemic dynamics. 2013; 2299–2302

241. Chen B, Hou G, Li A. Temporal local clustering coefficient uncovers the hidden pattern in temporal networks. Phys. Rev. E 2024; 109:064302

242. George B, Shekhar S. Time-aggregated graphs for modeling spatio-temporal networks. J. Data Semant. XI 2008; 191–212

243. Braha D, Bar-Yam Y. Time-dependent complex networks: Dynamic centrality, dynamic motifs, and cycles of social interactions. Adapt. Netw. Theory Models Appl. 2009; 39–50

244. Hadlak S, Schumann H, Cap CH, et al. Supporting the visual analysis of dynamic networks by clustering associated temporal attributes. IEEE Trans. Vis. Comput. Graph. 2013; 19:2267–2276

245. Pfitzner R, Scholtes I, Garas A, et al. Betweenness Preference: Quantifying Correlations in the Topological Dynamics of Temporal Networks. Phys. Rev. Lett. 2013; 110:198701

246. Scholtes I, Wider N, Garas A. Higher-order aggregate networks in the analysis of temporal networks: path structures and centralities. Eur. Phys. J. B 2016; 89:1–15

247. Cencetti G, Battiston F, Lepri B, et al. Temporal properties of higher-order interactions in social networks. Sci. Rep. 2021; 11:7028

248. Rohrschneider M, Ullrich A, Kerren A, et al. Visual network analysis of dynamic metabolic pathways. 2010; 316–327

249. Blonder B, Wey TW, Dornhaus A, et al. Temporal dynamics and network analysis. Methods Ecol. Evol. 2012; 3:958–972

250. Cardillo A, Petri G, Nicosia V, et al. Evolutionary dynamics of time-resolved social interactions. Phys. Rev. E 2014; 90:052825

251. Oettershagen L. Temporal graph algorithms. 2022;

252. Fründ J. Dissimilarity of species interaction networks: how to partition rewiring and species turnover components. Ecosphere 2021; 12:e03653

253. Sharma R, Kumar S, Song M. Fundamental gene network rewiring at the second order within and across mammalian systems. Bioinformatics 2021; 37:3293–3301

254. Boland MJ, Nazor KL, Loring JF. Epigenetic regulation of pluripotency and differentiation. Circ. Res. 2014; 115:311–324

255. Fernex D, Noack BR, Semaan R. Cluster-based network modeling—From snapshots to complex dynamical systems. Sci. Adv. 2021; 7:eabf5006

256. Zhang P, Gember-Jacobson A, Zuo Y, et al. Differential network analysis. 2022; 601–615

257. Lichtblau Y, Zimmermann K, Haldemann B, et al. Comparative assessment of differential network analysis methods. Brief. Bioinform. 2017; 18:837–850

258. Kivelä M, Arenas A, Barthelemy M, et al. Multilayer networks. J. Complex Netw. 2014; 2:203–271

259. Borgwardt KM. Graph kernels. 2007;

260. Haussler D. Convolution kernels on discrete structures. 1999;

261. Meng L, Striegel A, Milenković T. Local versus global biological network alignment. Bioinformatics 2016; 32:3155–3164

262. Milano M, Guzzi PH, Cannataro M. Glalign: A novel algorithm for local network alignment. IEEE/ACM Trans. Comput. Biol. Bioinform. 2018; 16:1958–1969

263. Kriege NM, Johansson FD, Morris C. A survey on graph kernels. Appl. Netw. Sci. 2020; 5:1–42

264. Guzzi PH, Milenković T. Survey of local and global biological network alignment: the need to reconcile the two sides of the same coin. Brief. Bioinform. 2018; 19:472–481

265. Singh R, Xu J, Berger B. Global alignment of multiple protein interaction networks with application to functional orthology detection. Proc. Natl. Acad. Sci. 2008; 105:12763–12768

266. De Domenico M, Lancichinetti A, Arenas A, et al. Identifying modular flows on multilayer networks reveals highly overlapping organization in interconnected systems. Phys. Rev. X 2015; 5:011027

267. Mucha PJ, Richardson T, Macon K, et al. Community structure in time-dependent, multiscale, and multiplex networks. science 2010; 328:876–878

268. Xu KS, Hero III AO. Dynamic stochastic blockmodels: Statistical models for time-evolving networks. 2013; 201–210

269. Xu KS, Hero AO. Dynamic stochastic blockmodels for time-evolving social networks. IEEE J. Sel. Top. Signal Process. 2014; 8:552–562

270. Afsarmanesh Tehrani N, Magnani M. Partial and overlapping community detection in multiplex social networks. 2018; 15–28

271. Palla G, Barabási A-L, Vicsek T. Quantifying social group evolution. Nature 2007; 446:664–667

272. Palla G, Derényi I, Farkas I, et al. Uncovering the overlapping community structure of complex networks in nature and society. nature 2005; 435:814–818

273. Han J, Cheng H, Xin D, et al. Frequent pattern mining: current status and future directions. Data Min. Knowl. Discov. 2007; 15:55–86

274. Berlingerio M, Pinelli F, Calabrese F. Abacus: frequent pattern mining-based community discovery in multidimensional networks. Data Min. Knowl. Discov. 2013; 27:294–320

275. Lin Y-R, Chi Y, Zhu S, et al. Facetnet: a framework for analyzing communities and their evolutions in dynamic networks. 2008; 685–694

276. Folino F, Pizzuti C. A multiobjective and evolutionary clustering method for dynamic networks. 2010; 256–263

277. Wang Y, Yao Y, Tong H, et al. A brief review of network embedding. Big Data Min. Anal. 2018; 2:35–47

278. Cui P, Wang X, Pei J, et al. A survey on network embedding. IEEE Trans. Knowl. Data Eng. 2018; 31:833–852

279. Grover A, Leskovec J. node2vec: Scalable feature learning for networks. 2016; 855–864

280. Perozzi B, Al-Rfou R, Skiena S. Deepwalk: Online learning of social representations. 2014; 701–710

281. Wang D, Cui P, Zhu W. Structural deep network embedding. 2016; 1225–1234

282. Nguyen GH, Lee JB, Rossi RA, et al. Continuous-time dynamic network embeddings. 2018; 969–976

283. Vatter J, Mayer R, Jacobsen H-A. The evolution of distributed systems for graph neural networks and their origin in graph processing and deep learning: A survey. ACM Comput. Surv. 2023; 56:1–37

284. Ma Y, Guo Z, Ren Z, et al. Streaming graph neural networks. 2020; 719–728

285. Qin M, Yeung D-Y. Temporal link prediction: A unified framework, taxonomy, and review. ACM Comput. Surv. 2023; 56:1–40

286. Gao C, Zheng Y, Li N, et al. A survey of graph neural networks for recommender systems: Challenges, methods, and directions. ACM Trans. Recomm. Syst. 2023; 1:1–51